\documentclass[a4paper,10pt]{article}
\usepackage{amssymb,amsmath,amsthm}
\usepackage{textcomp}

\usepackage{color}
\usepackage{latexsym}
\usepackage[english]{babel} 
\usepackage[latin1]{inputenc}  
\usepackage[T1]{fontenc}   
\usepackage[all]{xy}
\usepackage{epstopdf}
\usepackage{authblk}
\usepackage{stmaryrd}
\usepackage{amsfonts}
\usepackage{graphics}
\usepackage{epsfig}
\usepackage{graphicx}
\usepackage[colorlinks=true,linkcolor=blue]{hyperref}%
\usepackage{qcircuit}
\usepackage{braket}

\def\CC{\mathbb{C}}

\def\C{\mathbb{C}}

\def\PP{\mathbb{P}}

\usepackage{tikz}
\usetikzlibrary{automata}
\usetikzlibrary{shadows}
\usetikzlibrary{arrows}
\usetikzlibrary{shapes}
\usetikzlibrary{fit}
\usetikzlibrary{matrix}
\newcommand{\incl}{\ar@{^{}-}}
\newcommand{\inclu}{\ar@{^{}.}}

\newtheorem{proposition}{Proposition}[section]

\newtheorem{definition}{Definition  }[section]
\newtheorem{theorem}{Theorem}
\newtheorem{observation}{Observation}

\newtheorem{rem}{Remark    }[section]

\title{Grover's Algorithm and the Secant Varieties}
\author{Fr\'ed\'eric Holweck\footnote{frederic.holweck@utbm.fr}, Hamza Jaffali\footnote{hamza.jaffali@utbm.fr}, Isma\"el Nounouh\footnote{ismael.nounouh@utbm.fr}}
\affil{IRTES/UTBM,\\ Universit\'e de Bourgogne-Franche-Comt\'e, 90010 Belfort Cedex, France}
\begin{document}

\maketitle

\begin{abstract}
In this paper we investigate the entanglement nature of quantum states generated by Grover's search algorithm by means of algebraic geometry. More precisely we establish a 
link between entanglement of states generated by the algorithm and auxiliary algebraic varieties built from the set of separable states. 
This new perspective enables us to propose qualitative 
interpretations of earlier numerical results obtained by M. Rossi {\em et al}. We also illustrate our purpose with a couple of examples investigated in details.
\end{abstract}

\section{Introduction}
Grover's quantum search algorithm is a quantum algorithm which provides a quadratic speed-up when compared to the optimal 
classical search algorithms for unsorted database. 
When implemented on a multipartite quantum system ($n$-qudit), 
it generates an entangled state after its first iteration (the advantage of implementing Grover's algorithm on a multipartite
quantum system instead of a single $N$-dit Hilbert space 
is discussed by Meyer \cite{Meyer}).
The nature of this entanglement has been investigated numerically by various authors \cite{Chan, Fang, Wallach, Rossi1} by 
computing different measures of entanglement. 
For instance in the work of Rossi  {\em et al} \cite{Rossi1, Rossi2} one can find numerical computations of the  
Geometric Measure of Entanglement (GME) either as a function of the number of iterations for a fixed number of qubits \cite{Rossi1}
or as a function of the number of qubits when we only consider the first iteration of the algorithm \cite{Rossi2}. Those numerical approaches 
have the advantage to draw attention to the behavior of the algorithm and raise natural questions: when does the 
algorithm reach its 
maximum of entanglement ? How does it behave with several marked elements ?

In this note we will consider the same questions but from a different perspective, i.e. without any numerical approach.
We want to understand, in a more qualitative sense, which types of entangled states
are generated by the algorithm. More precisely using the geometric description of entanglement classes provided 
by auxiliary algebraic varieties (\cite{HLT}) we try to understand which stratas can be  reached (or not)
by the algorithm.

Let us recall some notations and a couple of definitions used in \cite{HLT}.
We consider $\mathcal{H}=\CC^{d_1}\otimes\CC^{d_2}\otimes\dots\otimes\CC^{d_m}$  the Hilbert space of states composed of $k$ particles, each being a $d_i$-dits. 
Denote by $|j_i\rangle$
 a basis of $\CC^{d_i}$ with $0\leq j_i\leq d_i-1$. A pure quantum  state $|\psi\rangle\in \mathcal{H}$ can be written as 
\[|\psi\rangle=\sum_{1\leq i\leq m}\sum_{0\leq j_{i}\leq d_{i}-1} a_{j_{1}j_{2}\dots j_{k}}|j_1\rangle\otimes\dots\otimes|j_m\rangle\]
 where $a_{j_1j_2\dots j_m}$ are complex amplitudes such that $\sum_{1\leq i\leq m}\sum_{0\leq j_{i}\leq d_{i}-1} |a_{j_1\dots j_m}|^2=1$,
 and $|j_1\rangle\otimes\dots\otimes|j_m\rangle$ is the standard basis of $\mathcal{H}$. 
 This basis will be denoted latter on by $|j_1\dots j_m\rangle$. 
 When $d_i=2~ \forall i$, i.e. $\mathcal{H}$ is a $n$-qubit Hilbert space, we will also 
 use the decimal notation for the basis, i.e the state $|j_1\dots j_m\rangle$ will be denoted by ${\bf \ket{x}}$ with 
 ${\bf x}=j_1.2^{m-1}+j_2.2^{m-2}+\dots+j_{m-1}2+j_m$.
Quantum states are uniquely determined up to a phase and the normalization factor does not provide meaningfull information. Therefore we can consider pure quantum  states $\ket{\psi}$ as points in 
the projectivized Hilbert space $[\psi]\in \PP(\mathcal{H})$. 
The complex semi-simple Lie group $G=SL(d_1,\CC)\times \dots\times SL(d_m,\CC)$ acts irreductibly on  $\mathcal{H}$ ($\mathcal{H}$ is a $G$-module).
The group $G$ is well-known in quantum information theory as the  group of (reversible) stochastic local quantum 
operations assisted
by classical communication (SLOCC \cite{Bennett,Miyake}). Under SLOCC  two states are equivalent if they are interconvertible by the action of $G$.


The $G$-module $\mathcal{H}$ has a unique highest weight vector which can be chosen to be $v=|0\dots0\rangle$ (it corresponds to a choice of orientation for the weight lattice \cite{FH}). 
The orbit $G.v\subset \mathcal{H}$ is the unique closed orbit for the action of $G$ on $\mathcal{H}$ and it
 defines, after projectivization, a smooth projective algebraic variety\footnote{In this paper a projective algebraic variety is understood as
 a subset $X\subset \PP(V)$ defined by the zero locus of a collection of homogeneous polynomials.}  $X=\PP(G.v)\subset \PP(\mathcal{H})$. 
This variety $X$ is known as the Segre embedding of the product
 of the projective spaces $\PP^{d_i-1}$, and it is the image of the  map \cite{Ha}:
\[\begin{array}{cccc}
   Seg: & \PP(\CC^{d_1})\times\PP(\CC^{d_2})\times\dots\times\PP(\CC^{d_m}) & \to & \PP(\CC^{d_1}\otimes\CC^{d_2}\otimes\dots\otimes\CC^{d_m})\\
             & ([v_1],[v_2],\dots,[v_m]) & \mapsto & [v_1\otimes v_2 \otimes \dots \otimes v_m] 
  \end{array}\]
where $v_i$ is a vector of $\CC^{d_i}$ and $[v_i]$ is the corresponding point in $\PP^{d_i-1}=\PP(\CC^{d_i})$.
The variety $X=\PP(G.v)=Seg(\PP(\CC^{d_1})\times\PP(\CC^{d_2})\times\dots\times\PP(\CC^{d_m}))$ will be simply denoted by
\[X=\PP^{d_1-1}\times\dots\times\PP^{d_m-1}\subset \PP(\mathcal{H})\] 

 From a quantum information theory point of view \cite{Brody,HLT,Hey}, the variety $X$ is the set of separable states in $\PP(\mathcal{H})$. Moreover if we suppose $d_i=2$ for all $i\in \llbracket 0,m\rrbracket$ then $X=\PP^1\times\dots\times \PP^1\subset \PP^{2^n-1}$ is 
 the variety of separable $n$-qubit.
 
The paper is organized as follow. In Section \ref{algo} we recall basic facts about Grover's algorithm and we make a usefull observation about the tensor rank of the states generated by the algorithm.
In Section \ref{secant} we use our observation to establish a first connection with auxiliary varieties. We show that for single marked element search, the algorithm always 
generates states which belong to the secant variety of the set of separable states. Our interpretation of the states generated by Grover's algorithm 
in terms of secant varieties leads us to a qualitative interpretation of the numerical computation of the GME proposed in \cite{Rossi1}. In particular we explain in Section \ref{Rossinum}
why the maximum of entanglement is obtained in \cite{Rossi1, Wallach} for specific values of $k$. We prove that, asymptotically, if $S$ is a set of orthogonal marked elements, the maximum of the GME is achieved
after $\frac{|S|}{|S|+1}k_{opt}$ iterations (Theorem \ref{maintheorem}) where $k_{opt}$ denotes the optimal number of iterations 
to be run before measurement. We also make a connection between the GME  of the quantum state generated after the first iteration 
as a function of the number of qubits as calculated in \cite{Rossi2} and the relative dimension of the corresponding auxiliary variety involved 
in our description. Finaly in Section \ref{tripartite} and Appendix \ref{A}, we describe explicitly all types of entangled 
classes reached by Grover's algorithm in geometrical terms  for single and multiple marked elements search in the
 $2\times 2\times 2$, $2\times 2\times 3$ and $2\times 3\times 3$ systems. Section \ref{conclusion} is dedicated to concluding remarks.
\section{Grover algorithm and tensor rank}\label{algo}
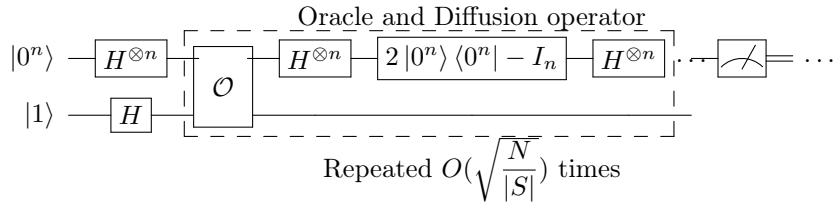
\begin{figure}[!h]
\[\Qcircuit @C=1em @R=.7em {
                &           &               &     & \mbox{Oracle and Diffusion operator}   &     &                   &\\
   \lstick{\ket{0^n}}& \gate{H^{\otimes n}}  & \multigate{1}{\mathcal{O}} & \gate{H^{\otimes n}} & \gate{2\ket{0^n}\bra{0^n}-I_n} & \gate{H^{\otimes n}} & \dots& \meter& \cw & \dots \\
   \lstick{\ket{1}} & \gate{H}  & \ghost{U^\dag}&\qw &\qw &\qw &\qw  \gategroup{2}{3}{3}{6}{.7em}{--}&\\
   & & &&&&&\\
    &                &        &           &        \mbox{{Repeated $O(\sqrt{\dfrac{N}{|S|}})$ times}} & &&
      }\]
  \caption{Grover's algorithm as a circuit}\label{circuit}
  \end{figure}

We first recall the principle of Grover's algorithm \cite{Grover,Lavor} when implemented on a $n$-qubit system. The algorithm starts with a $n$-qubit state whose registers are initialy on state $|0\rangle$, i.e. the initial state is $|\psi\rangle=|{\bf 0}\rangle=
{ |0\dots 0\rangle}$.
Employing a Hadamard gate on each register $H^{\otimes n}=\dfrac{1}{\sqrt{2}}\begin{pmatrix}
                                                                1 & 1\\
                                                                1 & -1
                                                               \end{pmatrix}^{\otimes n}$ one obtains the state corresponding to the superposition of all states of the computational basis  
                                                               $|\psi_0\rangle=\dfrac{1}{\sqrt{2^n}}\sum_{\text{x}=0} ^{2^n-1} |{\bf x}\rangle$.
                                                               Then the algorithm operates iteratively the so-called Grover gate $\mathcal{G}$ which is composed of two gates, the oracle $\mathcal{O}$ and the diffusion $\mathcal{D}$:
\begin{itemize}
\item The oracle corresponds to the unitary operator $\mathcal{O}={\bf 1}-2\sum_{{\bf x}\in S} |{\bf x}\rangle\langle {\bf x}|$ where $S$ is the set of elements in the computational basis which are sought and ``recognized'' by the oracle. 
When applied on a $n$-qubit state $|\psi\rangle=\sum_{{\bf x}=0} ^{2^n-1} \alpha_{\bf x} |{\bf x}\rangle$, 
the $\mathcal{O}$ gate signs the searched elements, $\mathcal{O}|\psi\rangle=-\sum_{{\bf x}\in S} \alpha_{\bf x} |{\bf x}\rangle+\sum_{{\bf x}\notin S} \alpha_{\bf x} |{\bf x}\rangle$. 
\item The diffustion gate $\mathcal{D}$ can be written as a unitary operator as $\mathcal{D}=-({\bf 1}-2|\psi_0\rangle\langle\psi_0|)$. This gate is also called inversion about the mean operation, it can be checked that 
$\mathcal{D}|\psi\rangle=\sum_{{\bf x}=0}^{2^n-1} (2\overline{\alpha}-\alpha_{\bf x})|{\bf x}\rangle$, where $\overline{\alpha}$ denotes the mean of the amplitudes $\alpha_{\bf x}$.
\end{itemize}
The algorithm can be encoded as a circuit (Figure \ref{circuit}).

After $k$-iterations the quantum state generated by Grover's algortihm is 
\begin{equation}\label{state}
 |\psi_k\rangle=\mathcal{G}^k|\psi_0\rangle= \dfrac{a_k}{\sqrt{|S|}}\sum_{{\bf x}\in S} |{\bf x}\rangle+ \dfrac{b_k}{\sqrt{2^{n}-|S|}}\sum_{{\bf x}\notin S} |{\bf x}\rangle
\end{equation}

with $a_k$ and $b_k$ real numbers such that $a_k^2+b_k^2=1$. Let $\theta$ be such that $\sin(\theta)=\sqrt{\dfrac{|S|}{2^{n}}}$, i.e. $\theta\approx \dfrac{\sqrt{|S|}}{2^{n-1}}$ 
for $|S|$ small compared to $2^{n}$. Then we can write (see \cite{Rieffel} page 182):
\begin{equation}\label{angle} a_k=\sin(\theta_k) \text{ and } b_k=\cos(\theta_k)\text{ with }\theta_k=(2k+1)\theta.\end{equation} This expression allows one to get the optimal number of iterations to get the highest possible probabilty to measure an element of $S$. Indeed 
the probability to obtain $|{\bf x}\rangle\in S$ after a measurement of $|\psi_k\rangle$ in the computational basis is $|a_k|^2=\sin(\theta_k)^2$. It will be optimal for $\theta_{k_{opt}}\approx \dfrac{\pi}{2}$, i.e. 
\begin{equation}\label{optimal} k_{opt}=\text{Round}[\dfrac{\pi}{4}\sqrt{\dfrac{2^{n}}{|S|}}] \end{equation} where 
$\text{Round}$ denotes the nearest integer function.

\begin{rem}
 Eq (\ref{optimal}) shows the quadratic speed up of the algorithm. For a database of $N=2^n$ elements, if there is only one element sought ($|S|=1$) then the complexity 
 of the algorithm is $O(\sqrt{N})$ compared to $O(\dfrac{N}{2})$ in all classical algorithms.
\end{rem}

Implemented on a multi-dits Hilbert space $\mathcal{H}=\CC^{d_1}\otimes \dots\otimes\CC^{d_m}$, the states $|\psi_k\rangle$ are tensors. When we deal with 
tensors one of the first attribute to consider is the rank \cite{Landsberg}. As pointed out in \cite{Bry},
tensor rank should be considered as an algebraic measure of entanglement.
\begin{definition}
 Let $\mathcal{H}$ be a Hilbert space obtained as tensor product of finite dimensional Hilbert spaces, i.e. $\mathcal{H}=\mathcal{H}_1\otimes\dots\otimes \mathcal{H}_m$ with $\text{dim }\mathcal{H}_i=d_i$. Then 
 $|\psi\rangle \in \mathcal{H}$ is said to be of 
 \begin{itemize} 
  \item rank $1$ if $|\psi\rangle=|u_1\rangle\otimes|u_2\rangle\otimes\dots\otimes|u_m\rangle$ with $|u_i\rangle\in \mathcal{H}_i$,
  \item rank $r$ if $|\psi\rangle=|\psi_1\rangle+\dots+|\psi_r\rangle$ where the $|\psi_i\rangle$ are rank $1$ tensors and $r$ is minimal with this property.
 \end{itemize}
\end{definition}

 It is clear from the definition that for pure muti-qudits system, rank one tensors correspond to separable states and every tensors which are not of rank one should be considered as entangled. Thus in order to study the entanglement generated by Grover's algorithm
 it is natural to ask what is the rank of the entangled states $|\psi_k\rangle$ of Eq (\ref{state}). The next observation will be essential in what follows.

 \begin{observation}\label{observation}
  If $S$ denotes the set of searched elements, then after $k$ iterations of the algorithm the state $\ket{\psi_k}$ can be written as
   \begin{equation}\label{tens}
   |\psi_k\rangle=\alpha_k\sum_{{\bf x}\in S} |{\bf x}\rangle+\beta_k|+\rangle^{\otimes n}
  \end{equation}
  where $\alpha_k, \beta_k$ are real numbers and $\ket{+}=\dfrac{1}{\sqrt{2}}(\ket{0}+\ket{1})$.
 \end{observation}

\proof The proof is straightforward. Consider the state $\ket{\psi_k}$ as given in Eq. (\ref{state}):
\begin{equation*}
 \begin{array}{lll}
  |\psi_k\rangle & = & \dfrac{a_k}{\sqrt{|S|}}\sum_{{\bf x}\in S} |{\bf x}\rangle+ \dfrac{b_k}{\sqrt{2^{n}-|S|}}\sum_{{\bf x}\notin S} |{\bf x}\rangle\\
                 & = & (\dfrac{a_k}{\sqrt{|S|}}-\dfrac{b_k}{\sqrt{2^{n}-|S|}})\sum_{{\bf x}\in S}\ket{{\bf x}}+\dfrac{b_k}{\sqrt{2^{n}-|S|}}\sum_{{\bf x}\in \{0,1\}^n} |{\bf x}\rangle\\
                 & = & \alpha_k\sum_{{\bf x}\in S}\ket{{\bf x}}+\beta_k \ket{+}^{\otimes n}
 \end{array}
\end{equation*}

with $\alpha_k=(\dfrac{a_k}{\sqrt{|S|}}-\dfrac{b_k}{\sqrt{2^{n}-|S|}})$ and $\beta_k=2^\frac{n}{2}\dfrac{b_k}{\sqrt{2^{n}-|S|}}$.

Observation \ref{observation} tells us in particular that the tensor rank of the states $\ket{\psi_k}$ genererated  by Grover's algorithm is bounded for $k<k_{opt}$:
\begin{equation}
 2\leq \text{Rank}(\ket{\psi_k})\leq |S|+1
\end{equation}

\begin{rem}
The upper bound is clear from Eq. (\ref{tens}). The lower bound is valid if  $\alpha_k\neq 0$ and $\beta_k\neq 0$. The fact that $\alpha_k\neq 0$ is insured by the convergence of the algorithm and if $\beta_k=0$ then 
$P(\ket{\psi_k} \in S)=1$, i.e.
$k=k_{opt}$.
\end{rem}

%
%
%
%
%
%

 \section{Grover's states as points of Secant varieties}\label{secant}
 It was first Heydari (\cite{Hey}) who pointed out the role of the secant varieties in describing classes of entanglement under SLOCC. This idea has been 
 investigated since then by various authors \cite{HLT, HLT2, HLT3, sanz, saw}.
 We recall its definition.
 \begin{definition}
  Let $X\subset \PP(V)$ be a projective algebraic variety. The secant variety of $X$ is the Zariski closure of secant lines of $X$:
  \begin{equation}
   \sigma(X)=\overline{\bigcup_{x,y\in X} \PP^1 _{xy}}.
  \end{equation}
 \end{definition}
 Higher order secant varieties may also be defined similarly:
 \begin{definition}
  Let $X\subset \PP(V)$ be a projective algebraic variety, then the $s^{\text{th}}$-secant variety of $X$ is the Zariski closure of secant 
  $(s-1)$-planes of $X$:
  \begin{equation}
   \sigma_s(X)=\overline{\bigcup_{x_1,\dots,x_s\in X} \PP^{s-1} _{x_1\dots x_s}}.
  \end{equation}
 \end{definition}
 In the case of $m$ distinguishable particles, i.e. $\mathcal{H}=\CC^{d_1}\otimes \dots \otimes \CC^{d_m}$, the (projective) variety of separable states $X=\PP^{d_1-1}\times\dots \times\PP^{d_k-1}$ 
is given by the Segre embedding. According to the Segre map any $[\psi] \in X$ is a rank one tensor and any rank one tensor is a point of $X$. It follows from the definition of $\sigma_s(X)$ that if 
$[\psi]$ is a general point of $\sigma_s(X)$ then there exists $[\psi_1],\dots,[\psi_s] \in X$ such that $[\psi]=[\psi_1+\dots+\psi_s]$, i.e. the general points of $\sigma_s(X)$ are tensors of rank  $s$.

\begin{definition}
Let $G$ a group acting on $\PP(V)$. A projective variety $Y\subset \PP(V)$ is a projective $G$-variety if $\forall y\in Y$ and $\forall g\in G$ we have $g.y\in Y$
\end{definition}
\begin{rem}
 By construction it is clear that if $X$ is a $G$-variety so are the secant varieties built from $X$. It will also 
 be true for other varieties obtained from $X$ by elementary geometric constructions. Such varieties are 
 called auxiliary varieties, see Section \ref{tripartite}.
\end{rem}

The variety of separable states being a SLOCC-orbit, it is clearly a SLOCC-variety and by construction so are the secant varieties.
More generally  if a pure state $[\psi]$ belongs to an auxiliary variety $Y$, all states SLOCC equivalent to $[\psi]$ will belong to the same variety $Y$. On the other hand if two pure states do not belong to the same auxiliary
variety we can conclude that the two states are not SLOCC equivalent.

The first secant variety has the nice property to be the orbit closure of the orbit of a general rank two tensor. Indeed if $[\psi] \in \sigma(X)=\sigma(\PP^{d_1}\times\dots \PP^{d_r})$ then there exist 
$[\psi_1]=[\ket{u_1\dots u_r}]$ and $[\psi_2]= [\ket{v_1\dots v_r}]\in X$, with $u_i, v_i\in \CC^{d_i}$, such 
that $[\psi]=[\psi_1+\psi_2]$ by definition of the secant variety. But then there exists $g=(g_1,\dots,g_r)\in GL_{d_1}\times \dots\times GL_{d_r}$ such that $g_i(u_i)=\ket{0}$ and $g_i(v_i)=\ket{1}$,
i.e. after projectivization there
exists $g\in$ SLOCC such that $g.[\psi]=[\ket{0}^{\otimes n}+\ket{1}^{\otimes n}]$. This last states is the well known generalized GHZ states.
\begin{equation}
\sigma(\PP^{d_1}\times\dots \times\PP^{d_r})=\overline{\text{SLOCC}.[\text{GHZ}_n]}.
\end{equation}
 The language of secant varieties allows us to state.
 \begin{proposition}\label{prop}
 \begin{enumerate}
  \item For one item searched, the states $[\psi_k]$ generated by Grover's algorithm for $0<k<k_{opt}$ are general points of the secant variety in particular the states $\ket{\psi_k}$  are $\text{SLOCC}$ equivalent to $\ket{\text{GHZ}_n}$.
  \item For multiple searched items, if the sought items $\ket{x_1},\dots, \ket{x_s} \in S$ are orthogonal and $|S|<<N=(d_1+1)\dots(d_r+1)$ then $[\psi_k]$ are general points of the $s+1$ secant variety.
 \end{enumerate}
 \end{proposition}
\proof It is a direct consequence of the Observation \ref{observation}. The searched elements $\ket{x_1}, \dots,\ket{x_s}$ of $S$ being orthogonal 
in the computationnal basis and $|S|<<N$, the $s$-dimensional plane $\PP^s_{[x_1],\dots,[x_s],[+^{\otimes n}]}$ is in general position 
in $\PP(\mathcal{H})$ and thus $[\psi_k]$ is a general point of $\sigma_{s+1}(X)$. $\Box$  \\
 
 This proposition offers a change of perspective in what was previously done to study the entanglement in Grover's algorithm. Instead of
 measuring a distance to the set of separable states (GME) we identify classes of entanglement with 
 specific SLOCC-varieties of the projectivized Hilbert space. This leads to qualitative interpretations of previous numerical computations 
 (Section \ref{Rossinum}) and 
 raises the question of  classification results about the types of entanglement which can be reached by 
 the algorithm (see Section \ref{tripartite} for first elementary examples).
\section{A geometric interpretation of the numerical results of {\em Rossi et al.}}\label{Rossinum}
The language of secant varieties and Proposition \ref{prop} offer new interpretations of the numerical results obtained by Rossi {\em et al.} in \cite{Rossi1,Rossi2}.
In \cite{Rossi1} the authors computed the Geometric Measure of Entanglement (GME) of states generated by Grover's algorithm for a $n$-qubit system.
\begin{equation}
 E_q(\ket{\psi})=1-\text{max}_{\phi\in \mathcal{S}_q} |\braket{\phi,\psi}|^2
\end{equation}

where $\mathcal{S}_q$ is the set of $q$-separable states, i.e. $\mathcal{S}_n$ in the variety of separable states and $\mathcal{S}_2$ is the set of the biseparable states.
The evolution of $E_q(\ket{\psi_k})$, as a function of $k$, is computed in $\cite{Rossi1}$ numerically for $n=12$ and $q\in\{2,n\}$ for a single searched item (case one of Proposition \ref{prop}) and two orthogonal searched items with $n=13$ and $q\in \{2,n\}$. 

\begin{figure}[!h]
\begin{center}
 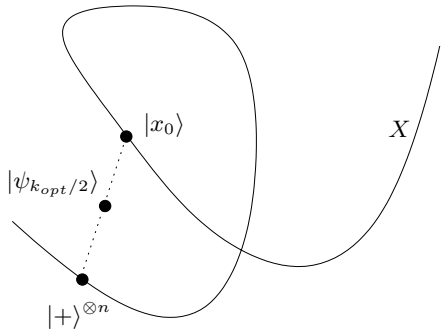
 \end{center}
 \caption{A pictural interpretation of the single searched item in Grover's algorithm evolution as point moving on a secant line. The ``curve'' $X$ represents the variety of separable states.}\label{seg}
\end{figure}

In the one searched item case, the evolution of the GME (both $q=2$ and $q=n$), as a function of the number of iterations $k$, starts from $0$ for $k=0$ and increases untill it 
reaches its maximum for $k\approx\dfrac{k_{opt}}{2}$ and then decreases up to $0$ for $k=k_{opt}$. 
We can point out that this result, encapsulated in  Figure 1 of \cite{Rossi1},  is qualitatively similar to the result of Meyer and Wallach (see Figure 1 of \cite{Wallach}).
The reason why the maximum of entanglement is reached at the middle of the algorithm is not explained in the paper.
Our Proposition \ref{prop} can be translated to a geometric picture, Figure \ref{seg}, which suggests why we could have expected this behavior:
If $\ket{\bf x_0}$ is the searched element, then the states $\ket{\psi_k}$ generated by  Grover's algorithm can be written as
\begin{equation}
 \ket{\psi_k}=\alpha_k \ket{\bf x_0}+\beta_k \ket{+}^{\otimes n}
\end{equation}

with $\alpha_k$ and $\beta_k$ are positive real numbers such that  for $k\in \llbracket 1,k_{opt}\rrbracket$,  $\alpha_k$ increases while $\beta_k$ decreases.
Therefore the states $\ket{\psi_k}$ evoluates during the algorithm on the secant line passing through the following two separable states $\ket{+}^{\otimes n}$ and $\ket{\bf x_0}$. At the beginning of the algorihm
we are in the initial states $\ket{\psi_0}=\ket{+}^{\otimes n}$ and when $k$ reaches $k_{opt}$ the states is close to $\ket{\bf x_0}$. It indicates that the maximum distance to the set of separable states should be achieved when 
$\ket{\psi_k}$ is close to the midpoint defined by $\ket{+}^{\otimes n}$ and $\ket{\bf x_0}$.

In the two searched (orthogonal) items case, the authors of \cite{Rossi1} perform a calculation for $n=13$ with $q=2$ and $q=n$. For the $q=n$ case the curve measuring the evolution of the GME with respect to $k$ increases from $0$ at $k=0$ untill it reaches 
a maximum for $k\approx \dfrac{2k_{opt}}{3}$ and then decreases to some nonzero value for $k=k_{opt}$ (See Figure 2 of \cite{Rossi1}). 
The reason why the GME is nonzero at the end of the algorithm is clear because when $k$ reaches $k_{opt}$ the state 
$[\psi_k]$ is close to be a point of the secant line $\PP^1_{\ket{\bf x_0},\ket{\bf x_1}}$ where $\ket{\bf x_0}$ and $\ket{\bf x_1}$ are the two orthogonal marked items. 
Thus $[\psi_{k_{opt}}]$ is not a point of $X$. The secant picture also suggests  a reason 
why the maximum of entanglement is obtained for $k\approx \dfrac{2k_{opt}}{3}$. As the state $[\psi_k]$ moves on the secant 
plane $\PP^2_{\ket{+}^{\otimes n},\ket{\bf x_0},\ket{\bf x_1}}$ from $[\ket{+}^{n}]$ to the midpoint of the segment
joining $[\ket{\bf x_0}]$ and $[\ket{\bf x_1}]$ we expect its maximum distance from the set of separable states to be achieved when $[\psi_k]$
is close to the barycenter.
\begin{figure}[!h]
\begin{center}
 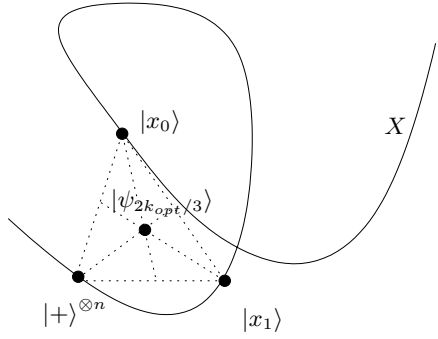
 \end{center}
 \caption{A pictural interpretation of the two orthogonal searched items in Grover's algorithm evolution as a point moving on a secant plane. The ``curve'' $X$ represents the variety of separable states.}\label{plane}
\end{figure}

This barycenter effect, suggested by Figures \ref{seg} and \ref{plane}, which explains qualitatively the numerical results of Rossi {\em et al.} \cite{Rossi1} can be more precisely stated.
\begin{theorem}\label{maintheorem}
 Let $\mathcal{H}=(\C^d)^{\otimes n}$ be the Hilbert space of a $n$ $d$-dit system. Let us denote by $S$ a set of  orthogonal marked elements with $|S|\leq d$. Then for $n$ large 
 the measure of entanglement achieves its maximum for $k\approx\dfrac{|S|}{|S|+1}k_{opt}$ with $k_{opt}=\text{Round}[\dfrac{\pi}{4}\sqrt{\dfrac{d^n}{|S|}}]$. 
\end{theorem}

\proof 
Let $S=\{\ket{x_1},\dots,\ket{x_{s}}\}$ be the set of orthogonal marked elements. In $\mathcal{H}$ we consider the convex hull $K$ defined by the states $\ket{+}^{\otimes n}, \ket{x_1},\dots,\ket{x_s}$.
\begin{equation}
 K=\mathcal{C}(\ket{+}^{\otimes n}, \ket{x_1},\dots,\ket{x_s}).
\end{equation}
The Grover state 
$\ket{\psi_k}$ moves from $\ket{+}^{\otimes n}$ towards the state $\ket{\psi}=\dfrac{1}{\sqrt{s}}(\ket{x_1}+\dots+\ket{x_s})$. 
\begin{equation}
 \ket{\psi_k}=\alpha_k\ket{\psi}+\beta_k\ket{+}^{\otimes n}.
\end{equation}

The distance
of $\ket{\psi_k}$ to the vertices of $K$ is maximal when $\ket{\psi_k}$ reaches a position close to  
the barycenter of $K$. Under the assumption $|S|<<d^n$ we have $\alpha_k\approx a_k$ and $\theta_k=(2k+1)\theta$ (Eq \ref{angle}). Thus $\ket{\psi_k}$ is 
close to the barycenter of $K$ when $\theta_k\approx \dfrac{s}{s+1}\dfrac{\pi}{2}$, therefore 
when $k\approx \dfrac{s}{s+1}k_{opt}$.
However the distance from $\ket{\psi_k}$ to the vertices of $K$ may not be equal to the distance to the set of separable states. 

Because of 
orthogonality and $s\leq d$ we can assume that the marked states $\ket{x_i}$ are all symmetric. For instance one can choose $\ket{x_1}=\ket{0}^{\otimes n}$, $\ket{x_2}=\ket{1}^{\otimes n}$, $\dots$, $\ket{x_s}=\ket{s-1}^{\otimes n}$.
The marked states being symmetric we have $\ket{\psi_k}$ is symmetric. Thus according to \cite{Hub} the measure of entanglement 
can be obtained by restricting to symmetric separable states.
\begin{equation}
 E_n(\psi_k)=1-\text{max}_{\phi \in X, \phi \text{ symmetric}} |\langle \psi_k,\phi\rangle|^2.
\end{equation}

Let $\ket{\phi}=(\delta_1 \ket{0}+\dots+\delta_p\ket{p-1})^{\otimes n}$ be a symmetric separable states with $p\leq d$. 
Then for $\ket{\psi_k}= \alpha_k (\ket{0}^{\otimes n}+\dots+\ket{s-1}^{\otimes n})+\beta_k\ket{+}^{\otimes n}$ we get 
\begin{equation}
 |\langle \psi_k,\phi\rangle|^2=|\alpha_k(\delta_1^n+\dots+\delta_s^n)+\beta_k(\dfrac{\delta_1+\dots+\delta_p}{\sqrt{p}})^n|^2.
\end{equation}

If we denote by $m=\text{max}(|\beta_k|,|\alpha_k|)$ we obtain 

\begin{equation}
 |\langle \psi_k,\phi\rangle|^2\leq m^2|(\delta_1^n+\dots+\delta_s^n+\dfrac{\delta_1+\dots+\delta_p}{\sqrt{p}})^n|^2.
\end{equation}

We can assume the $\delta_i$ to be positive number and also $0\leq \delta_i\leq 1$ and $0\leq \dfrac{\delta_1+\dots+\delta_p}{\sqrt{p}}\leq 1$.

Let us look for the maximum of \begin{equation} f(\delta_1,\dots,\delta_p)=\delta_1^n+\dots+\delta_s^n+(\dfrac{\delta_1+\dots+\delta_p}{\sqrt{p}})^n\end{equation}

For $n$ large each terms $\delta_i ^n$ imposes the existence of a local maximum in the neigborhood of $\delta_i=1$, $\delta_j=0$ for $j\neq i$ and similarly
the term $(\dfrac{\delta_1+\dots+\delta_p}{\sqrt{p}})^n$ imposes the existence of a local maximum in the neigborhood of 
$\delta_1=\dots=\delta_p=\dfrac{1}{\sqrt{p}}$.
But $\delta_i=1\pm \varepsilon$ leads to $\ket{\phi}=\ket{i-1}^{\otimes n}+\varepsilon \ket{\tilde{\phi}}$ and $\delta_1=\dots=\delta_p=\dfrac{1}{\sqrt{p}}$ 
corresponds to $\ket{\phi}=\ket{+}^{\otimes n}+\varepsilon\ket{\tilde{\phi}}$.
In other words we obtain for $n$ large
\begin{equation}
\begin{array}{lll}
 E_n(\psi_k)& = & 1-\text{max}_{\phi \in X, \phi \text{ symmetric}} |\langle \psi_k,\phi\rangle|^2\\
            & = & 1-\text{max}_{\phi \in X, \phi \in K} |\langle \psi_k,\phi\rangle|^2 +\mathcal{O}(\varepsilon)\\
            & \approx & 1-\text{max}_{\phi \in X, \phi \in K} |\langle \psi_k,\phi\rangle|^2.
 \end{array}
\end{equation}
Thus for $n$ large we can restrict the calculation of $E_n$ to an optimization on the vertices of $K$. The maximum of $E_n$ 
is therefore obtained when $\ket{\psi_k}$ is close to the barycenter of $K$. $\Box$\\

There is an other numerical calculation of Rossi {\em et al.} proposed in \cite{Rossi2} which can be given a geometric explanation 
based on the interpretation in terms of secant varieties.
In \cite{Rossi2} the authors calculated for $n$-qubit system the value $E_n(\psi_1)$, i.e. the Geometric Measure of 
Entanglement after the first iteration, as a function of $n$ the number of qubits for one and two marked elements. Their results are given 
in Figure 1 of \cite{Rossi2}. For the different cases under consideration, the corresponding curves show the same behavior, 
i.e. an exponential decay.
From our perspective for one or two marked elements the state $\ket{\psi_1}$ is a general point of the first or second secant variety, i.e.
$\sigma(X)$ or $\sigma_3(X)$. But the dimensions 
of those varieties increase linearly as a function of $n$  while the dimension of the ambient space increases exponentially. More precisely it is known \cite{Landsberg} that 
\begin{equation}
 \text{dim}(\sigma_k(X))\leq k\text{dim}(X)+k-1
\end{equation}
 with equality in the general case. In particular for $X=\underbrace{\PP^1\times\dots \times \PP^1}_{n \text{ times}}$ we have for $n>2$.
 
 \begin{equation}
  \text{dim}(\sigma(X))=2n+1.
 \end{equation}

\begin{figure}[!h]
 \begin{center}
  \includegraphics[width=6cm]{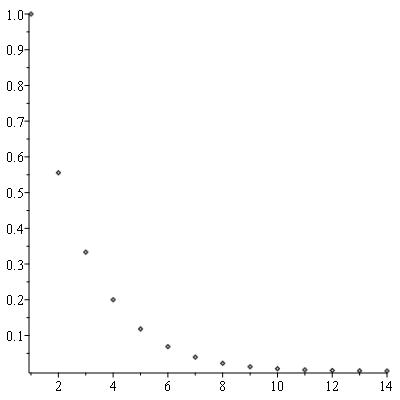}
  \caption{Normalized relative dimension of the first secant variety as a function of the number of qubits (to be compared with Figure 1 of \cite{Rossi2})}\label{relsec}
 \end{center}

\end{figure}

 The relative dimension of the first secant variety compared to the dimension of the ambient space is therefore given by 
 \begin{equation}
  RD_{\sigma}:n\mapsto \dfrac{2n+1}{2^n-1}.
 \end{equation}
If we normalize this function such that it is equal to $1$ for $n=1$, we obtain the normalized relative dimension 
of the first secant variety\begin{equation} NRD_{\sigma}(n)=\dfrac{1}{3}(\dfrac{2n+1}{2^n-1})\end{equation}
The behavior of the curve of the normalized relative dimension of the first secant variety
 (Figure \ref{relsec}) is similar  to the plotting of the GME of $\ket{\psi_1}$ as a function of $n$ given in  \cite{Rossi2}.

The similarity of those two curves (Figure \ref{relsec} of the present paper and Figure 1 of \cite{Rossi2}) can be understood as follow.
For one marked element the first state $\ket{\psi_1}$ generated by Grover's algorithm is  always a general point 
of the first secant variety (Proposition \ref{prop}) and the GME measures the distance of this point to the variety of separable states. But it is a relative 
distance in the sense that the GME is always bounded by $1$.
As shown by Figure \ref{relsec}, as the dimension of the ambient space increases, the relative dimension of 
the first secant variety decreases
exponentially. The GME is maximal for points which are general points of  the ambient space.
Thus the (relative) distance of $\ket{\psi_1}$ to the set of separable states decreases at the same rate as the relative dimension of the first secant variety.

More interesting is the case of two marked elements. For two marked elements the authors of \cite{Rossi2} have computed the GME of $\ket{\psi_1}$ as a function of 
the number of qubits for different configurations of marked elements i.e. with marked elements having Hamming distance $1,2,3$ or $4$. Because the Hamming distance is not maximum the states under consideration 
are not generic points of $\sigma_3(\PP^1\times\dots \times\PP^1)$. For instance when the Hamming distance is one, the sum of the two marked elements is a separable state and thus 
the states $\ket{\psi_1}$ belongs to the first secant variety. However no matter in which variety the state $\ket{\psi_1}$ is, the relative dimension
of the variety will again decrease exponentially because 
\begin{equation}\label{dimsec}
 \text{dim}(\sigma_3(\PP^1\times\dots\times\PP^1))\leq 3n+2.
\end{equation}
\begin{rem}
 The GME values of $\ket{\psi_1}$ of the four cases under consideration (Hamming distance $1,2, 3$ or $4$) in \cite{Rossi2} are very close except in the 
 $n=4$ case. For $n=4$, the Hamming distance equal to $4$ corresponds to orthogonal marked states and therefore the states $\ket{\psi_1}$ is a general point of  $\sigma_3(\PP^1\times \PP^1\times\PP^1\times\PP^1)$
 while the ones corresponding to marked elements with Hamming distance $1,2$ or $3$ will be points of subvarieties of the third secant variety. It is interesting to point out that 
 for $n$-qubit systems, the case $n=4$ is the only one where the inequality of Eq (\ref{dimsec}) is not an equality (see \cite{Geramita}). The stratification of the four-qubit 
 Hilbert space exhibits a rich structure (\cite{HLT2, HLP}) which could explain the different values obtained in  \cite{Rossi2} in this case.
\end{rem}

\section{Examples from  tripartite quantum systems}\label{tripartite}

 In this section we  calculate for some tripartite systems   which SLOCC classes are reached by states generated by Grover's algorithm. The cases we consider are the $2\times 2\times 2$, the $2\times 2\times 3$ and the 
 $2\times 3\times 3$ quantum systems.
 We focus on those cases because the number of orbits is finite (and thus the SLOCC classification is complete). Moreover a geometric description of those orbits in terms of auxiliary varieties as well as invariants/covariants polynomials to identify them
 have been given in \cite{HLT}. Thus for any given state of those systems we can tell by the results of \cite{HLT} in which auxiliary varieties the state belong.
 
 \subsection{The number of marked elements}
 To prove the quadratic speed-up of the algorithm it is assumed that $|S|$, the number of marked elements, is small compared to the dimension of the Hilbert space $|S|<<N$. It is also one of the 
 assumption of Theorem \ref{maintheorem}. In fact from a pratical point of view we should assume $|S|<\dfrac{N}{4}$ to obtain at least one iteration before we reach the optimal state. 
 This situation will be called {\em standard case}.
 
 The case where $|S|=\dfrac{N}{4}$ is pecular and 
 we will call it {\em critical case}. In this case if we consider the initial state $\ket{\psi_0}=\ket{\bf 0}=\sum_{{\bf x}\in S} \ket{{\bf x}}+\sum_{{\bf x}\notin S} \ket{{\bf x}}$ then 
 the first iteration of the Grover gate $\mathcal{G}$ leads to $\mathcal{G}\ket{\psi_0}=\sum_{{\bf x}\in S}\ket{{\bf x}}$, i.e. the only state reached by the algorithm is the sum of the marked elements. 
 The optimal state is obtained after the first iteration
 and we see that every states which are sum of $|S|$ elements in the computational basis will define  a SLOCC orbit reachable by the algorithm.
 
 The cases where $|S|>\dfrac{N}{4}$ is less interesting but can still be computed, it will be called {\em exceptional case}
 
 Therefore in the next calculations we will always consider the three different types of regime:
 
 \begin{enumerate}
  \item $|S|<\dfrac{N}{4}$ the standard regime: the natural situation to apply Grover's algorithm.
  \item $|S|=\dfrac{N}{4}$ the critical case.
  \item $|S|>\dfrac{N}{4}$ the exceptional case.
 \end{enumerate}

 \subsection{The join of two varieties}
To describe geometrically the SLOCC-stratas of the tripartite systems considered in this section we will need to define the join of two varieties $X$ and $Y$.
The join of two projective varieties $X$ and $Y$ is defined by
 \begin{equation}\label{join}
  J(X,Y)=\overline{\bigcup_{x\in X,y\in Y} \PP_{xy}^1}.
 \end{equation}
According to Eq (\ref{join}), the $s^\text{th}$-secant variety can be described inductively as a sequence of join  varieties.
\begin{equation}
\sigma_s(X)=J(X,\sigma_{s-1}(X)).
\end{equation}
If $Y\subset X$ we denote by $T^* _{X,Y,y_0}$ the union of lines $\PP^1_*$ where $\PP_*^1$ is the limit of the lines $\PP^1_{xy}$ with $x\in X, y\in Y$ and $x,y\to y_0$. 
The union of $T^*_{X,Y,y_0}$ is named by Zak  \cite{Zak} the variety of tangent stars of $X$ with respect to $Y$:
\begin{equation}
 T(Y,X)=\cup_{y\in Y} T_{X,Y,y} ^*.
\end{equation}

\begin{rem}
 If $X$ is smooth and $Y=X$, the variety $T(X,X)$ is nothing but the union of all tangent lines to $X$, i.e. the tangential variety, usually denoted by $\tau(X)$.
\end{rem}

The tensor product structure of $\mathcal{H}=\CC^{d_1}\otimes \dots \otimes \CC^{d_m}$ allows one to introduce classes of sub-secant varieties.

\begin{definition}
  Let $X\subset \PP(\mathcal{H})$ be a projective algebraic variety, and let $J=\{j_1,\dots,j_p\}\subset \{1,\dots,m\}$ then the $s^{\text{th}}$-$J$-secant variety of $X$ is the Zariski closure of secant
  $(s-1)$-planes of $X$:
  \begin{equation}
   \sigma_s ^J(X)=\overline{\bigcup_{x_1,\dots,x_s\in X} \PP^{s-1} _{x_1\dots x_s}}
  \end{equation}
  where the $x_i$'s satisfy the following conditions,
  $x_i=v_1 ^i\otimes \dots \otimes u_{j_1}\otimes\dots v_l ^i \otimes \dots \otimes u_{j_p} \otimes \dots \otimes v_m ^i$, with $v_j ^i \in \CC^{d_j}$ and $u_{j_t} \in \CC^{d_{j_t}}$.
 \end{definition}
 
\subsection{The 3 qubits case}
The SLOCC classification of orbits in $\mathcal{H}=\CC^2\otimes \CC^2\otimes \CC^2$ is known in the quantum information community since the work of D\"ur, Vidal and Cirac \cite{Dur} but an explicit 
list of normal forms and the Hasse diagram of the orbit closure can be found in the work of Parfenov \cite{Pav} or in the book \cite{GKZ}. To avoid normalization, the orbits, which are conical, are considered 
in the projective Hilbert space $\PP^7=\PP(\mathcal{H})$ (thus the trivial orbit is omitted).
We reproduce in Table \ref{222} the six SLOCC orbits of this classification with, for each orbit, a normal form 
and the description of \cite{HLT} in terms of algebraic variety of the orbit closure.
\begin{table}[!h]
\begin{center}
\begin{tabular}{|c|c|c|c|}
\hline
Orbit  & Normal form (representative) & Variety (orbit closure) & Dimension\\
\hline
${\mathcal{O}}_{6}$ & $|000\rangle+|111\rangle $&  $\PP^7$ & $7$  \\ 
${\mathcal{O}}_5$ & $|100\rangle+|010\rangle+|001\rangle$& $\tau(\PP^1\times \PP^1 \times \PP^1)$& $6$ \\
${\mathcal{O}}_{4}$ & $|001\rangle+|111\rangle$  &   $\sigma(\PP^1\times \PP^1)\times \PP^1$& $4$\\ 
${\mathcal{O}}_{3}$& $|100\rangle+|111\rangle$ & $\PP^1\times\sigma(\PP^1\times \PP^1)$ & $4$ \\
${\mathcal{O}}_{2}$ & $|010\rangle+|111\rangle$ & $\sigma(\PP^1\times\underline{\PP^1}\times \PP^1)\times \PP^1$ & $4$ \\
${\mathcal{O}}_1$ & $|000\rangle$ &  $\PP^1\times \PP^1 \times \PP^1$ & $3$\\
\hline
\end{tabular}
\caption{Identification of orbit closures and varieties for the $2\times 2\times 2$ quatum system}\label{222}
\end{center}
\end{table}


We recoginze the so-called $\ket{GHZ}$ and $\ket{W}$ states as normal forms of the $\mathcal{O}_6$ and the $\mathcal{O}_5$ orbit. 
Their orbits form a dense subset of respectively the
secant variety and the tangential variety. In particular one sees that the $\ket{W}$ state is in the closure of the orbit of the $\ket{GHZ}$ state. It is a well known 
fact in algebraic geometry which has been restated and interpreted in the language of quantum information theory in \cite{HL,sanz}.

\begin{figure}
\begin{equation}
 \xymatrix{&  \mathcal{O}_6&  \\ 
 &\mathcal{O}_5\incl[u] &  \\
     \mathcal{O}_2 \incl[ru] \incl[rd] & \mathcal{O}_3\incl[u] \incl[d]& \mathcal{O}_4 \incl[ul]  \incl[dl]\\
  & \mathcal{O}_1&  \\
  }
  \end{equation}
\caption{Hasse diagram of the orbit closures}\label{hass}
\end{figure}
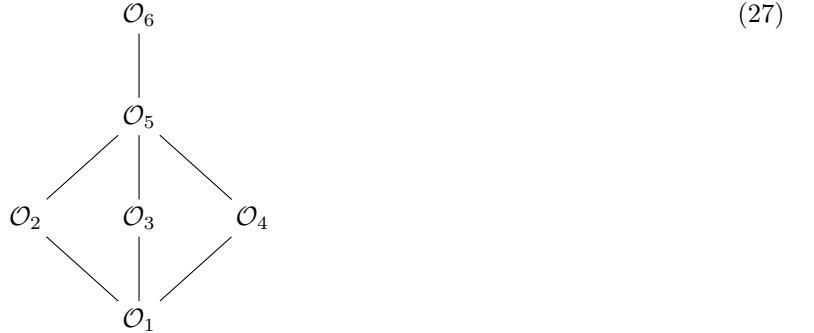

The hierachy between the orbit closures can be sketched by a Hasse diagram (Figure \ref{hass}).

Using the techniques of \cite{HLT} to identify a states as point of an orbit we can show that,
\begin{itemize}
 \item For $|S|=1$ the states generated by Grover's algorithm belong to $\mathcal{O}_6$ as expected by Proposition \ref{prop}.
 \item For $|S|=2$ (critical case) the states generated by Grover's algorithm belong to $\mathcal{O}_1$, $\mathcal{O}_2$, $\mathcal{O}_3$
 and $\mathcal{O}_4$. This is  not a surprise because 
 for the critical case  the states generated by the algorithm are the sum of the marked elements. But all normal forms of the orbits $\mathcal{O}_1, \mathcal{O}_2, \mathcal{O}_3$ and $\mathcal{O}_4$ can 
 be written as sum of two basis state. It is clear for the orbits $\mathcal{O}_2$, $\mathcal{O}_3$ and $\mathcal{O}_4$ but it is also true for $\mathcal{O}_1$ 
 because $\ket{000}+\ket{001}=\ket{00}\ket{+}\in \mathcal{O}_1$.
 \item For $|S|>2$ (by symmetry we may assume $|S|\leq 4$), 
 the orbits $\mathcal{O}_1$, $\mathcal{O}_2$, $\mathcal{O}_3$ and $\mathcal{O}_4$ can be obtained.
\end{itemize}

Table \ref{222res} in  Appendix \ref{A} provide an example for the orbit $\mathcal{O}_6$ of marked elements which will generate
states in that orbit in the $|S|<\dfrac{N}{4}$ mode.

 We point out that the orbit corresponding to $\ket{W}$ is not reached by the states generated by the algorithm.
 
\begin{figure}[!h]
 \begin{center}
  \includegraphics[width=7cm]{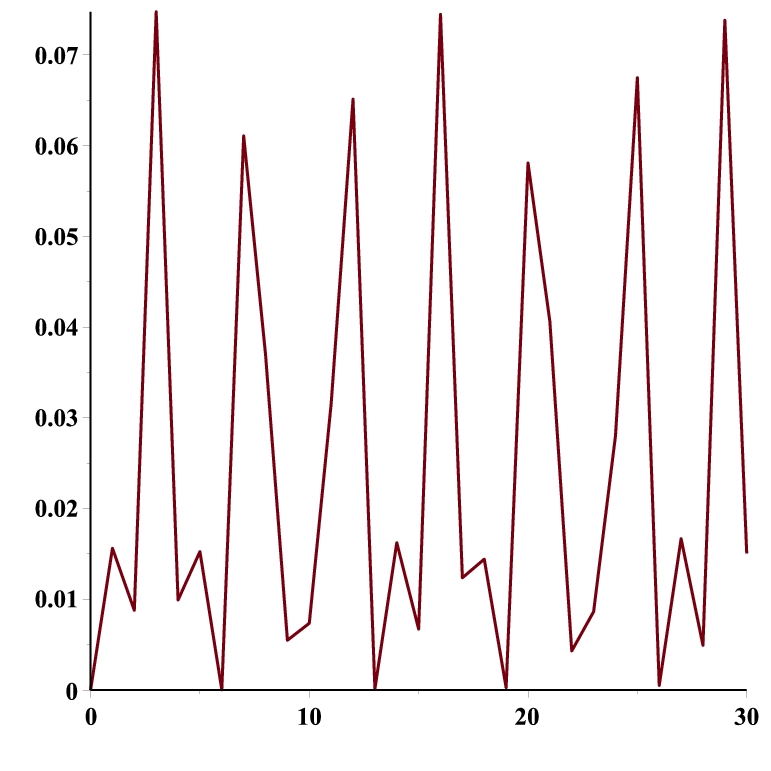}
  \caption{Evolution of $k\mapsto |\Delta_{222}(\ket{\psi_k})|$ for the set of marked elements $S=\{\ket{000}\}$}\label{d222}
  \end{center}
 \end{figure}

 The polynomial defining the orbit closure of $\ket{W}$  is known as the Cayley (or $2\times 2\times 2$-)-hyperdeterminant \cite{GKZ}, 
 $\Delta_{222}$. It is the unique (up to scale) invariant 
 polynomial of degree $4$ for the algebra of three qubits and its module can be used as a measure of entanglement (it is nothing but the square of the $3$-tangle).
 When we plot the evolution of $|\Delta_{222}(\psi_k)|$ for the one search item, for example $S=\{\ket{000}\}$, as a function of $k$, we recover the periodical behavior of the algorithm (Figure \ref{d222}).

\subsection{The $2\times 2\times 3$ case}
In this case there are $8$ non trivial SLOCC orbits to consider (Table \ref{223}). 
\begin{table}[!h]
\begin{center}
\begin{tabular}{|c|c|c|c|}
\hline
Orbit  &Normal form  & Variety (orbit closure) & Dimension\\
\hline
 ${\mathcal{O}}_{8}$&$|000\rangle+|011\rangle+|101\rangle+|112\rangle$&       $\PP^{11}$ &$11$\\
${\mathcal{O}}_{7}$ &$|000\rangle+|011\rangle+|102\rangle$& $J(X,\overline{\mathcal{O}}_{IV})$ &$10$ \\
${\mathcal{O}}_{6}$ &$|000\rangle+|111\rangle$& $\sigma(X)$ & $9$\\
${\mathcal{O}}_{5}$& $|000\rangle+|011\rangle+|101\rangle$& $\tau(X)$ &$8$ \\
${\mathcal{O}}_{4}$&$|000\rangle+|011\rangle$&  $\PP^1\times\sigma(\PP^1\times \PP^2)\simeq\PP^1\times \PP^{5}$ & $6$\\
${\mathcal{O}}_{3}$&$|000\rangle+|101\rangle$   & $\sigma(\PP^1\times \underline{\PP^1}\times \PP^2)\times \PP^1$ &$6$\\
${\mathcal{O}}_{2}$&$|000\rangle+|110\rangle$&$\sigma(\PP^1\times \PP^1)\times \PP^2\simeq \PP^3\times \PP^2$ &$5$\\
${\mathcal{O}}_{1}$&$|000\rangle$& $X=\PP^1\times \PP^1 \times \PP^2$  &$4$\\
\hline
 \end{tabular}
 \caption{Identification of orbit closures and varieties for the $2\times 2\times 3$ quantum system}\label{223}
\end{center}
\end{table}
The dimension of the Hilbert space $\mathcal{H}=\CC^2\times\CC^2\times \CC^3$ being equal to $12$ we will 
consider Grover's algorithm with the number of marked elements being $|S|\leq 2$. Under this constrain we obtain.
\begin{itemize}
 \item $|S|=1$ the orbit $\mathcal{O}_6$ is reached (Proposition \ref{prop}).
\item $|S|=2$ the orbits $\mathcal{O}_3$, $\mathcal{O}_4$, $\mathcal{O}_7$ and $\mathcal{O}_8$ are reached.
 \end{itemize}

\begin{figure}[!h]
 \begin{center}
  \includegraphics[width=7cm]{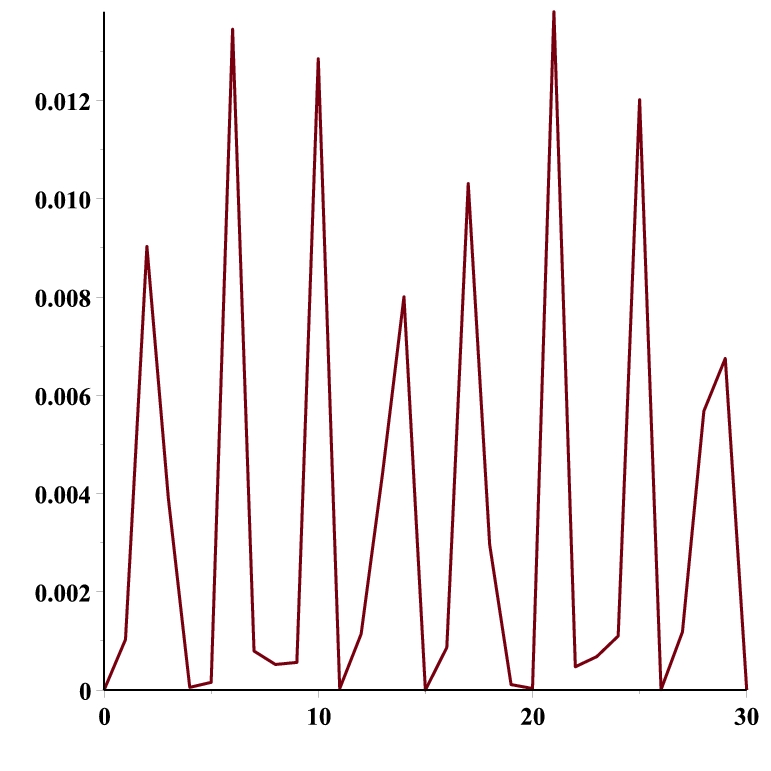}
  \caption{Evolution of $k\mapsto |\Delta_{223}(\ket{\psi_k})|$ for the set of marked elements $\{\ket{000},\ket{111}\}$}\label{d223}
  \end{center}
 \end{figure}

In the critical case $|S|=3$ all states  can be reached by the algorithm except the orbit $\mathcal{O}_8$ and thus 
for $|S|>3$ no new orbits are obtained. 
Again if we disregard the critical case ($|S|=3$) one  notice that the orbit which corresponds to the natural 
generalization of the $\ket{W}$ state (orbit $\mathcal{O}_5$)  is not reached by the algorithm.
Like in the three qubits case the defining equation of the hypersurface $J(X,\overline{\mathcal{O}}_4)$ is an invariant polynomial which we denote by 
$\Delta_{223}$ and is named as the $2\times 2\times 3$ hyperdeterminant. It is the generator of the 
ring of SLOCC invariant polynomials for the  $2\times 2\times 3$ system. Its module can be used as a measure of entanglement 
and we can plot the curve $k\mapsto |\Delta_{223}(\psi_k)|$ when 
$\ket{\psi_k}$ belongs to $\mathcal{O}_8$.
We reproduce the curve obtained for two marked elements $S=\{\ket{000},\ket{111}\}$ (Figure \ref{d223}).
 
\begin{figure}[!h]
 \begin{center}
  \includegraphics[width=7cm]{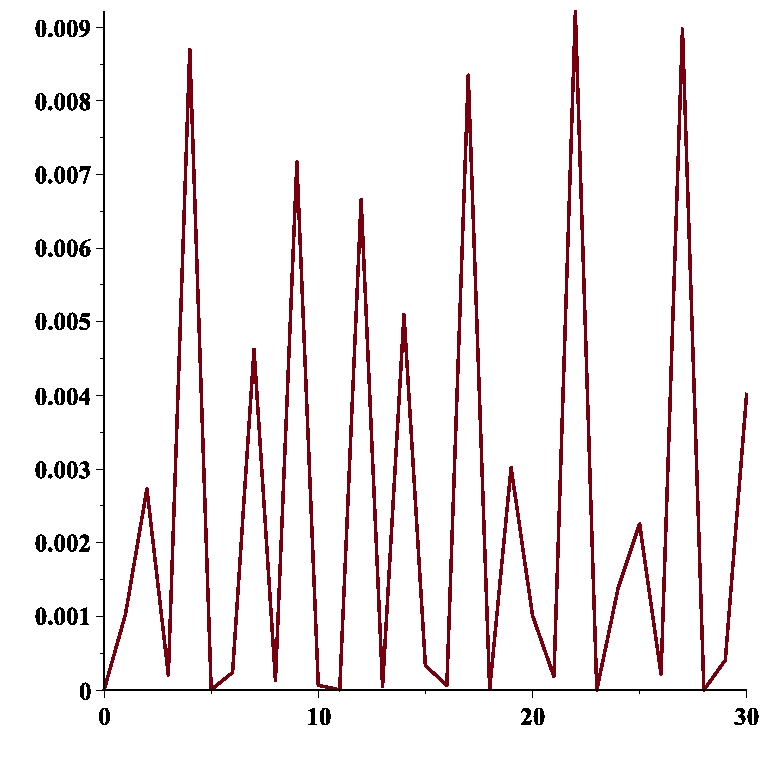}
  \caption{Evolution of $k\mapsto |\Delta_{223}(\ket{\psi_k})|$ for the set of marked elements $\{\ket{000},\ket{100},\ket{010},\ket{001}\}$}\label{d223bis}
  \end{center}
 \end{figure}

As shown in Appendix \ref{A} Table \ref{223res} there are different ways of choising a set of marked elements which will generate states in $\mathcal{O}_8$.
For instance if we choose $S=\{\ket{000},\ket{100},\ket{010},\ket{001}\}$, then the states $\ket{\psi_k}$ also belong to $\mathcal{O}_8$. We can then 
plot the alternative curve $k\mapsto |\Delta_{223}(\psi_k)|$ for this new choice of $S$ (Figure \ref{d223bis}).

\begin{figure}[!h]
 \begin{center}
  \includegraphics[width=7cm]{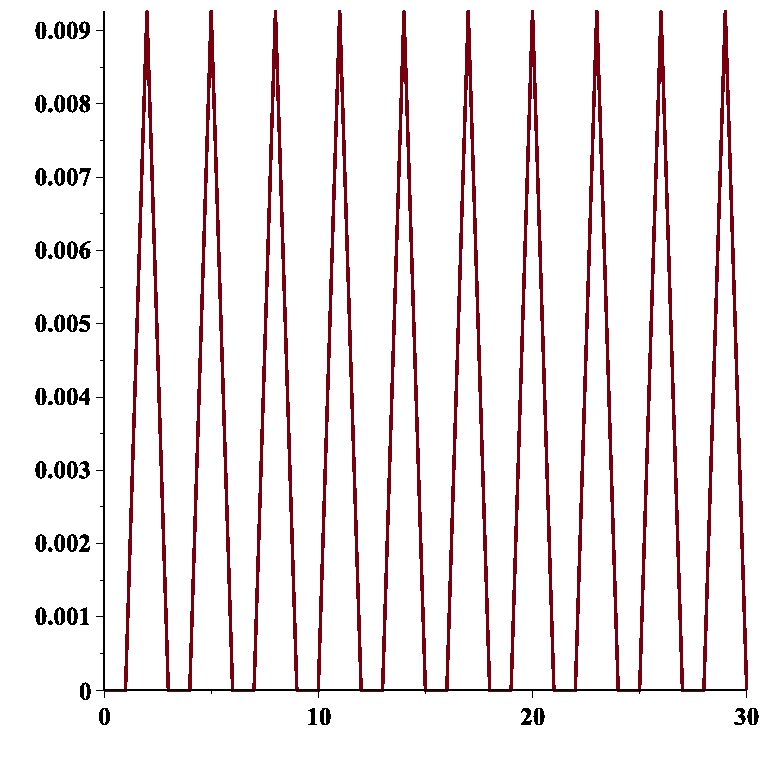}
  \caption{Evolution of $k\mapsto |\Delta_{223}(\ket{\psi_k})|$ for the set of marked elements $\{\ket{000},\ket{110},\ket{101}\}$}\label{d223tri}
  \end{center}
 \end{figure}

Finally we can also plot the curve in the critical case for $S=\{\ket{000},\ket{110},\ket{101}\}$ (Figure \ref{d223tri}). 
One sees that the behavior of $k\mapsto |\Delta_{223}(\psi_k)|$ is really pecular 
in the critical case.
 
 Table \ref{223res} of Appendix \ref{A} also illustrates the importance of the multipartite structure of the Hilbert space in consideration. 
 For instance for
  two marked elements, depending on the choice of the marked elements, the algorithm generates states which do not belong to 
 the same SLOCC-orbit: For instance if $S=\{\ket{000},\ket{101}\}$, we obtain a Grover state $\ket{\psi_k}$ which is a general point 
 of $\mathcal{O}_7$ but if $S=\{\ket{000},\ket{110}\}$, one obtains a Grover state $\ket{\psi_k}$ which is a general point of $\mathcal{O}_6$.
 
\subsection{The $2\times 3\times 3$ case}
In this last case the orbit structure is richer as there are $17$ SLOCC-orbits (Table \ref{233}).
\begin{table}[!h]
\begin{center}
 \begin{tabular}{|c|c|c|c|}
\hline
Orbit  &Normal form & Variety (orbit closure) & Dimension\\
\hline
${\mathcal{O}}_{17}$&$|000\rangle+|011\rangle+|100\rangle+|122\rangle$& $\PP^{17}$ &$17$ \\
${\mathcal{O}}_{16}$& $|000\rangle+|011\rangle+|101\rangle+|122\rangle$&        $J(X,\tau(X))$ &$16$\\
${\mathcal{O}}_{15}$&$|000\rangle+|011\rangle+|022\rangle+|101\rangle+|112\rangle$& $T(X,\tau(X))$ &$15$ \\
${\mathcal{O}}_{14}$&$|000\rangle+|011\rangle+|122\rangle$& $J(X,\PP^1\times\sigma(\PP^2\times\PP^2))$ & $14$\\
${\mathcal{O}}_{13}$&$|000\rangle+|011\rangle+|022\rangle+|101\rangle$& $T(X,\PP^1\times\sigma(\PP^2\times\PP^2))$ &$13$ \\
${\mathcal{O}}_{12}$&$|000\rangle+|011\rangle+|101\rangle+|112\rangle$&  $\sigma(\sigma(\PP^1\times\underline{\PP^2}\times\PP^2)\times\PP^2)$ & $13$\\
${\mathcal{O}}_{11}$&$|000\rangle+|011\rangle+|121\rangle+|102\rangle$   & $J(\PP^5\times\PP^2,\sigma(\PP^1\times\underline{\PP^2}\times\PP^2)\times\PP^2)$ &$13$\\
${\mathcal{O}}_{10}$&$|000\rangle+|011\rangle+|102\rangle$& $J(X,\sigma(\PP^1\times\underline{\PP^2}\times\PP^2)\times\PP^2))$ &$12$\\
${\mathcal{O}}_{9}$&$|000\rangle+|011\rangle+|022\rangle$& $\PP^1\times\sigma_3(\PP^2\times\PP^2)\simeq \PP^1\times\PP^8$  &$9$\\
${\mathcal{O}}_{8}$&$|000\rangle+|011\rangle+|110\rangle+|121\rangle$& $\sigma(\PP^5\times\PP^2)$  &$13$\\
${\mathcal{O}}_{7}$&$|000\rangle+|011\rangle+|120\rangle$& $J(X,\PP^5\times\PP^2)$  &$12$\\
${\mathcal{O}}_{6}$&$|000\rangle+|111\rangle$& $\sigma(X)$  &$11$\\
${\mathcal{O}}_{5}$&$|000\rangle+|011\rangle+|101\rangle$& $\tau(X)$  &$10$\\
${\mathcal{O}}_{4}$&$|000\rangle+|011\rangle$& $\PP^1\times\sigma(\PP^2\times\PP^2)$  &$8$\\
${\mathcal{O}}_{3}$&$|000\rangle+|101\rangle$& $\sigma(\PP^1\times\underline{\PP^2}\times\PP^2)\times\PP^2$  &$7$\\
${\mathcal{O}}_{2}$&$|000\rangle+|110\rangle$& $\sigma(\PP^1\times\PP^2)\times\PP^2\simeq \PP^5\times\PP^2$  &$7$\\
${\mathcal{O}}_{1}$ &$|000\rangle$& $X=\PP^1\times\PP^2\times\PP^2$  &$5$\\
\hline
 \end{tabular}
\caption{Identification of orbit closures and varieties for $2\times 3\times 3$ quantum system}\label{233}

\end{center}
\end{table}

If we consider $|S|\leq 4$ all orbits can be reached by the algorithm except the orbits $\mathcal{O}_5$, $\mathcal{O}_{11}$, $\mathcal{O}_{13}$ and $\mathcal{O}_{15}$.
More precisely we have:
\begin{itemize}
 \item For $|S|=1$ only the orbit $\mathcal{O}_6$ (the secant variety) is obtained (Proposition \ref{prop}).
 \item For $|S|=2$, the orbits $\mathcal{O}_4, \mathcal{O}_6, \mathcal{O}_7, \mathcal{O}_{10}, \mathcal{O}_{14}$ and $\mathcal{O}_{17}$ can be reached.
 \item For $|S|=3$, the algorithm generates states of the orbits $\mathcal{O}_2, \mathcal{O}_3, \mathcal{O}_6, \mathcal{O}_7, \mathcal{O}_8, \mathcal{O}_{10}, \mathcal{O}_{12},\mathcal{O}_{14}, \mathcal{O}_{16}$ and $\mathcal{O}_{17}$ 
\item For $|S|=4$, one can obtain the orbits $\mathcal{O}_4, \mathcal{O}_6, \mathcal{O}_7, \mathcal{O}_8, \mathcal{O}_9, \mathcal{O}_{10},  \mathcal{O}_{12}, \mathcal{O}_{14}, \mathcal{O}_{16}$ and $\mathcal{O}_{17}$
 \end{itemize}

 For this system there is no critical case, thus if we allow $|S|\geq 5$ we find that the orbit $\mathcal{O}_{11}$ and $\mathcal{O}_{13}$ can also be obtained by Grover's algorithm. However the orbits $\mathcal{O}_{5}$
 and $\mathcal{O}_{15}$
 can never be obtained as states generated by the algorithm. If we look at the geometric intepretation given by Table \ref{233} of the closures
 of those orbits, one sees that 
 they all correspond to tangential varieties, i.e. tensors which are limits of joins of the variety of separable states. If we only consider the standard regime, $|S|<\dfrac{N}{4}$, no tangential varieties, including the variety corresponding to the orbit closure of the analogue of the $\ket{W}$-state (orbit $\mathcal{O}_5$), can be reached by the algorithm.
 
 It will be interesting to check if that would always be the case. For instance if we limit ourselves to qubits can it be proven that the states $\ket{W}_n=\ket{10\dots0}+\ket{01\dots0}+\dots+\ket{00\dots1}$ is never produced by the algorithm
 except in the critical situation $|S|=\dfrac{N}{4}$?
 
\begin{figure}[!h]
 \begin{center}
  \includegraphics[width=7cm]{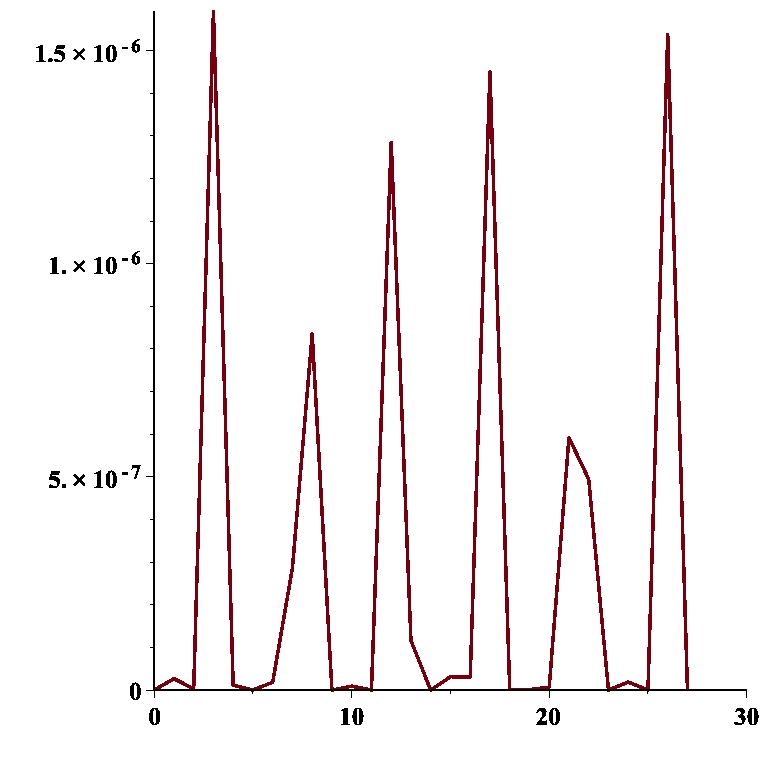}
  \caption{Evolution of $k\mapsto |\Delta_{233}(\ket{\psi_k})|$ for the set of marked elements $\{\ket{000},\ket{111}\}$}\label{233_1}
  \end{center}
 \end{figure}

\begin{figure}[!h]
 \begin{center}
  \includegraphics[width=7cm]{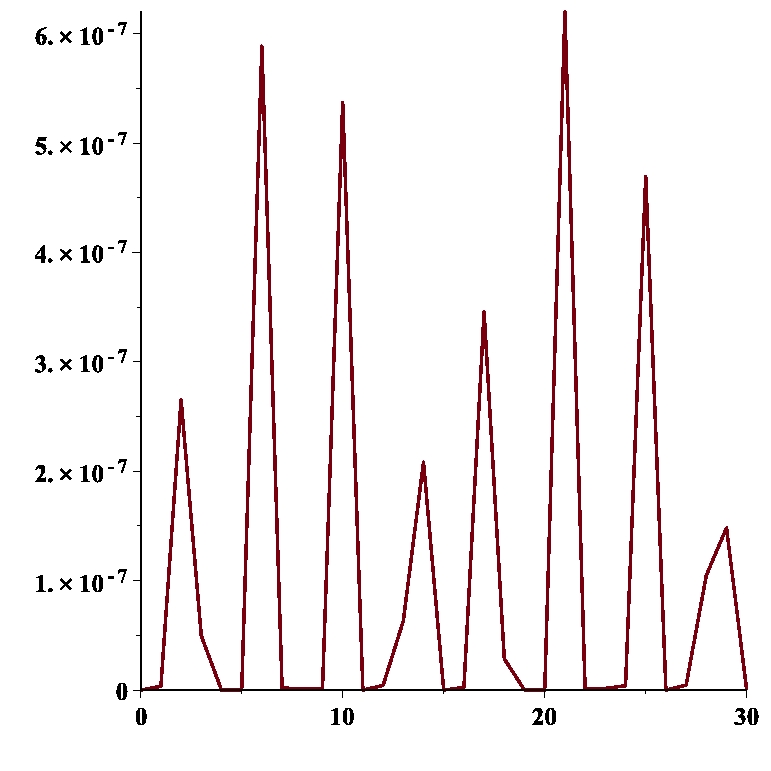}
  \caption{Evolution of $k\mapsto |\Delta_{233}(\ket{\psi_k})|$ for the set of marked elements $\{\ket{000},\ket{001}, \ket{110}\}$}\label{233_2}
  \end{center}
 \end{figure}

\begin{figure}[!h]
 \begin{center}
  \includegraphics[width=7cm]{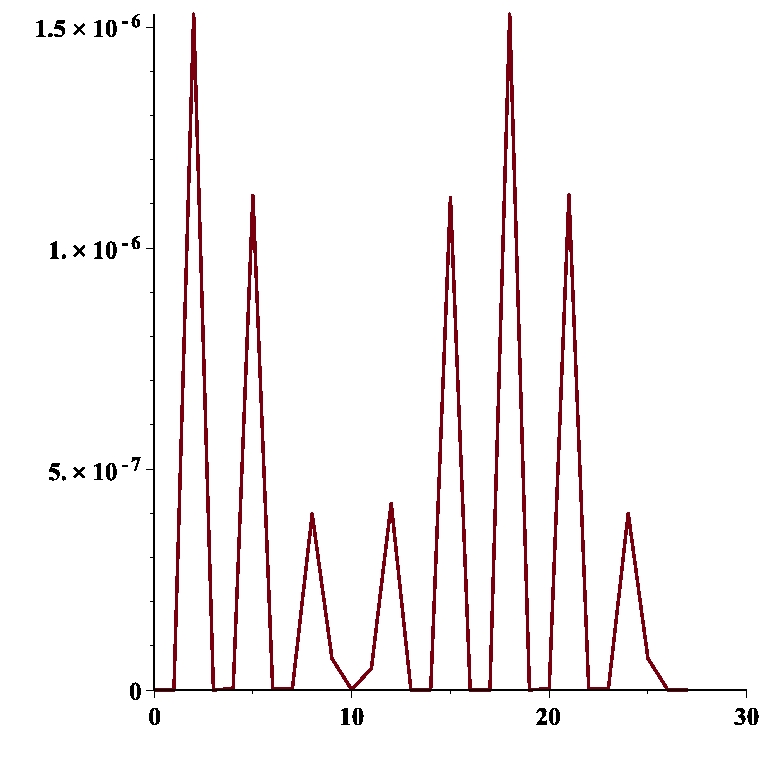}
  \caption{Evolution of $k\mapsto |\Delta_{233}(\ket{\psi_k})|$ for the set of marked elements $\{\ket{000},\ket{001}, \ket{010},\ket{102}\}$}\label{233_3}
  \end{center}
 \end{figure}

 In this case the dense orbit $\mathcal{O}_{17}$ can be obtained in the standard regime with two, three or four marked elements. 
 If we plot the variation, as a function of $k$, of 
 the module of the $2\times 3\times 3$-hyperdeterminant $k\mapsto |\Delta_{233}(\psi_k)|$ one obtains three different curves illustrating again 
 the periodicity  of the algorithm (Figures \ref{233_1}, \ref{233_2}, \ref{233_3}).
 
 In Apprendix \ref{A} Table \ref{233res}, we provide examples of choice of marked elements and the corresponding orbits 
 obtained  when applying Grover's algorithm to  those set of marked elements. Like in the $2\times 2\times 3$ we clearly see the influence of the implementation on 
 a multipartite system.

\section{Conclusion}\label{conclusion}
In this paper we propose a new qualitative investigation of the nature of entanglement generated by Grover's algorithm. By employing the language of secant and auxiliary varieties we provided 
geometrical explanations 
of numerical results obtained by Rossi {\em et al} \cite{Rossi1,Rossi2}. This geometrical perspective confirms the numerical results and anticiaptes on further possible calculations. 
If we think about the entanglement classes as (open subset of) SLOCC invariant algebraic varieties, our calculation also leads 
to the more general question: which entangled classes can be obtained by Grover's algorithm? By working on a few examples one showed that some specific classes, which share the same geometric interpretation 
are not reachable by states generated by the algorithm. It is in particular the case for the $\ket{W}$ state and its generalization in the $2\times 2\times 3$ and $2\times 3\times 3$ Hilbert 
spaces.
The next case which can be worked out by our techniques is the case of four-qubit states. 
In this case the number of orbits is infinite but there are $9$ families (some depending on parameters, see \cite{V}) which 
allow to describe all possibles orbits and there exists an algorithm based on invariant/covariant to identify a given state with its SLOCC-equivalent family up to a qubit permutation \cite{HLT3}. 

\subsection*{Acknowledgment}
The authors would like to thank Prof. Jean-Gabriel Luque for kindly providing them his Maple code to compute the invariants/covariants used in the 
calculation of Section \ref{tripartite}.

\appendix
\section{Examples of marked elements}\label{A}
The following Tables provide examples of sets of marked elements which allows to reach the corresponding orbits by running Grover's algorithm in the standard regime $|S|<\dfrac{N}{4}$.
\begin{table}[!h]
 \begin{center}
  \begin{tabular}{|c|c|}
  \hline
   Orbit & $|S|=1$  \\
   \hline
  $\mathcal{O}_6$ & $\{\ket{000}\}$    \\
  $\mathcal{O}_5$ & --- \\
  $\mathcal{O}_4$ & --- \\
  $\mathcal{O}_3$ & --- \\
  $\mathcal{O}_2$ & --- \\
  $\mathcal{O}_1$ & ---  \\  
  \hline
  \end{tabular}
\caption{Examples of family of marked elements $S$ and the corresponding orbits reached by the algorithm in the $2\times 2\times 2$ case.}\label{222res}
 \end{center}

\end{table}
\begin{table}[!h]
 \begin{center}
  \begin{tabular}{|c|c|c|}
  \hline
   Orbit & $|S|=1$ & $|S|=2$  \\
   \hline
  $\mathcal{O}_8$ &  ---  &$\{\ket{000},\ket{111}\}$\\
  $\mathcal{O}_7$ &  ---  &$\{\ket{000},\ket{101}\}$\\  
  $\mathcal{O}_6$ & $\{\ket{000}\}$  &  $\{\ket{000},\ket{110}\}  $ \\
  $\mathcal{O}_5$ &  ---&     ---                        \\
  $\mathcal{O}_4$ &  --- &    $\{\ket{000},\ket{100}\}$                     \\
  $\mathcal{O}_3$ &  --- &  $ \{\ket{000},\ket{010}\} $                     \\
  $\mathcal{O}_2$ &  --- &   ---                       \\
  $\mathcal{O}_1$ &  --- &   ---                        \\  
  \hline
  \end{tabular}
\caption{Examples of family of marked elements $S$ and the corresponding orbits reached by the algorithm in the $2\times 2\times 3$ case.}\label{223res}
 \end{center}

\end{table}

\begin{table}[!h]
 \begin{center}
  \begin{tabular}{|c|c|c|c|c|}
  \hline
   Orbit & $|S|=1$ & $|S|=2$ & $|S|=3$ & $|S|=4$ \\
   \hline
    
  $\mathcal{O}_{17}$ &--- &          $ \{\ket{000},\ket{111}\} $               &   $  \{\ket{000},\ket{001},\ket{110}\}   $        & $\{\ket{000},\ket{001},\ket{010},\ket{102}\}$\\
  $\mathcal{O}_{16}$ & --- &         ---                &        $ \{\ket{000},\ket{011},\ket{101}\} $      &$\{\ket{000},\ket{001},\ket{010},\ket{100}\} $\\
  $\mathcal{O}_{15}$ &---  &        ---                 &    ---         &---\\
  $\mathcal{O}_{14}$ & --- &                 $ \{\ket{000},\ket{011}\} $      &      $ \{\ket{000},\ket{001},\ket{010}\}  $    &$\{\ket{000},\ket{001},\ket{010},\ket{012}\}$\\
  $\mathcal{O}_{13}$ & --- &      ---                   &    ---         &--- \\  
    $\mathcal{O}_{12}$ &---  &    ---   &    $ \{\ket{000},\ket{010},\ket{121}\}  $  &     $\{\ket{000},\ket{001},\ket{110},\ket{120}\}  $   \\
  $\mathcal{O}_{11}$ &--- &     ---                      &      ---          &--- \\
  $\mathcal{O}_{10}$ &---  &         $ \{\ket{000},\ket{101}\}     $          &   $  \{\ket{000},\ket{001},\ket{100}\}$           &$ \{\ket{000},\ket{001},\ket{010},\ket{020}\}$\\
  $\mathcal{O}_9$ &---  &     ---                    &     ---        &$\{\ket{000},\ket{011},\ket{100},\ket{111}\}$\\
  $\mathcal{O}_8$ & --- &     ---                    &        $ \{\ket{000},\ket{001},\ket{112}\} $   &$\{\ket{000},\ket{001},\ket{012},\ket{102}\}$\\
  $\mathcal{O}_7$ & --- &              $ \{\ket{000},\ket{110}\} $         &       $\{\ket{000},\ket{001},\ket{012}\}$      & $\{\ket{000},\ket{001},\ket{002},\ket{010}\}$\\  
  $\mathcal{O}_6$ &$\{\ket{000}\}$ &   $\{\ket{000},\ket{001}\}$    &     $\{\ket{000},\ket{001},\ket{102}\}$    &   $\{\ket{000},\ket{001},\ket{002},\ket{100}\} $      \\
  $\mathcal{O}_5$ &--- &      ---                     &   ---             &--- \\
  $\mathcal{O}_4$ &---  &          $  \{\ket{000},\ket{100}\}  $           &        ---        &$ \{\ket{000},\ket{001},\ket{100},\ket{101}\}$\\
  $\mathcal{O}_3$ & --- &     ---                    &    $  \{\ket{000},\ket{010},\ket{020}\} $      &---\\
  $\mathcal{O}_2$ & --- &     ---                    &    $ \{\ket{000},\ket{001},\ket{002}\}    $    &---\\
  $\mathcal{O}_1$ & --- &     ---                    &       ---      & ---\\  
  \hline
  \end{tabular}
\caption{Examples of family of marked elements $S$ and the corresponding orbits reached by the algorithm in the $2\times 3\times 3$ case.}\label{233res}
 \end{center}

\end{table}

\end{document}